\documentclass[12pt, leqno]{article}
\title{On various relations with FRLW cosmology}
\author{Jennie D'Ambroise}

\usepackage{amssymb, amsmath, textcomp}
\usepackage{graphicx}
\usepackage{epsfig}     
\usepackage{dsfont,upgreek}
\usepackage{amsthm}
\usepackage{mathrsfs}
\usepackage{mathtools, verbatim}
\usepackage[all]{xy}

\newtheoremstyle{dotless}{}{}{\itshape}{}{\bfseries}{}{ }{}
\theoremstyle{dotless}

\numberwithin{equation}{section}

\begin{document}
 \begin{flushleft}
      \Large \bf
      \strut {Applications of Elliptic and Theta Functions to 
   Friedmann-Robertson-Lema\^itre-Walker Cosmology with Cosmological Constant}
      \par
    \end{flushleft}
\begin{center}
      \Large \bf
      \strut Jennie D'Ambroise
\end{center}
\section{\protect Introduction}
Elliptic functions are known to appear in many problems, applied and theoretical.  However, a lesser known application is in the study of exact solutions to Einstein's gravitational field equations in a Friedmann-Robertson-Lema\^itre-Walker (FRLW) cosmology \cite{ACB, AurichSteiner, AurichSteinerThen, B-HFB, Kharb, KW}.   We will show explicitly how Jacobi and Weierstrass elliptic functions arise in this context, and will additionally show connections with theta functions.  In section 2, we review the definitions of various elliptic functions.  In section 3, we record relations between elliptic functions and theta functions.   In section 4 we introduce the FRLW cosmological model and we then proceed to show how elliptic functions appear as solutions to Einstein's gravitational equations in sections 5 and 6.  The author thanks Floyd Williams for helpful discussions. 

\section{\protect Elliptic functions}

\indent\indent An \emph{elliptic integral} is one of the form $\int R(x,\sqrt{P(x)})dx$ where $P(x)$ is a polynomial in $x$ of degree three or four and $R$ is a rational function of its arguments.  Such integrals are called elliptic since an integral of this kind arises in the computation of the arclength of an ellipse.  Legendre showed that any elliptic integral can be written in terms of the three fundamental or \emph{normal elliptic integrals}
\begin{equation*}F(x,k)\stackrel{def.}{=}\displaystyle\int_0^x\frac{dt}{\sqrt{(1-t^2)(1-k^2t^2)}}\end{equation*}
\begin{equation}E(x,k)\stackrel{def.}{=}\displaystyle\int_0^x\sqrt{\frac{1-k^2t^2}{1-t^2}}dt\end{equation}
\begin{equation*}\Pi(x,\alpha^2, k)\stackrel{def.}{=}\displaystyle\int_0^x\frac{dt}{(1-\alpha^2t^2)\sqrt{(1-t^2)(1-k^2t^2)}}\end{equation*}
which are referred to as normal elliptic integrals of the first, second and third kind, respectively.  The parameter $\alpha$ is any real number and $k$  is referred to as the \emph{modulus}.  For many problems in which real quantities are desired, $0<k,x< 1$, although this is not required in the above definitions (for $k=0$ and $k=1$, the integral can be expressed in terms of elementary functions and is therefore \emph{pseudo-elliptic}).

\emph{Elliptic functions} are inverse functions of elliptic integrals.  They are known to be the simplest of non-elementary functions and have applications in the study of classical equations of motion of various systems in physics including the pendulum.  One can easily show that if $f(u)$ denotes the inverse function of an elliptic integral $y(x)=\displaystyle\int R(x,\sqrt{P(x)})dx$, then since $y(f(u))=u \Rightarrow y'(f(u))=1/f'(u)$, 
\begin{equation}f'(u)^2=\frac{1}{R(f(u),\sqrt{P(f(u))})^2}.\end{equation}

The \emph{Jacobi elliptic function}  $sn(u,k)$ is the inverse of $F(x,k)$ defined above, and eleven other Jacobi elliptic functions can be written in terms of $sn(u,k)$:  $cn(u,k)$ and $dn(u,k)$ satisfy
\begin{equation}sn^2u+cn^2u=1\mbox{ and }k^2sn^2u+dn^2u=1\end{equation}
respectively, and 
\begin{eqnarray}ns\stackrel{def.}{=}\frac{1}{sn},&nc\stackrel{def.}{=}\mbox{\Large{$\frac{1}{cn}$}},&nd\stackrel{def.}{=}\frac{1}{dn}\notag\\
sc\stackrel{def.}{=}\frac{sn}{cn},& sd\stackrel{def.}{=}\mbox{\Large{$\frac{sn}{dn}$}}, & cd\stackrel{def.}{=}\frac{cn}{dn}\\
cs\stackrel{def.}{=}\frac{1}{sc},& ds\stackrel{def.}{=}\mbox{\Large{$\frac{1}{sd}$}}, & dc\stackrel{def.}{=}\frac{1}{cd}.\notag\end{eqnarray}  
By (2.1), (2.2), (2.3) and (2.4) one can see that each one of the Jacobi elliptic functions satisfy the differential equation
\begin{equation}f'(u)^2+af(u)^2+bf(u)^4=c\end{equation}
 for some $a,b,c$ in terms of the modulus $k$ according to the following table.
\begin{center}
\begin{tabular}{|c|c|c|c|}
\hline
$f(u)$&$a$&$b$&$c$\\
\hline
$sn(u,k)$&$1+k^2$&$-k^2$&$1$\\
$cn(u,k)$&$1-2k^2$&$k^2$&$1-k^2$\\
$dn(u,k)$&$k^2-2$&$1$&$k^2-1$\\
$ns(u,k)$&$1+k^2$&$-1$&$k^2$\\
$nc(u,k)$&$1-2k^2$&$k^2-1$&$-k^2$\\
$nd(u,k)$&$k^2-2$&$1-k^2$&$-1$\\
$sc(u,k)$&$k^2-2$&$k^2-1$&$1$\\
$sd(u,k)$&$1-2k^2$&$k^2(1-k^2)$&$1$\\
$cd(u,k)$&$1+k^2$&$-k^2$&$1$\\
$cs(u,k)$&$k^2-2$&$-1$&$1-k^2$\\
$ds(u,k)$&$1-2k^2$&$-1$&$k^2(k^2-1)$\\
$dc(u,k)$&$1+k^2$&$-1$&$k^2$\\
\hline
\end{tabular}
\end{center}
Note that the $a$-value for $f(u)=dn(u,k)$ seen above corrects an error in \cite{B-HFB}.

\indent\indent The \emph{Weierstrass elliptic function} 
\begin{equation}\wp(z;\omega_1,\omega_2)=\frac{1}{z^2}+\hspace{-.15in}\displaystyle\sum_{(m,n)\in\mathds{Z}\times\mathds{Z}-\{(0,0)\}}
\frac{1}{(z-m\omega_1-n\omega_2)^2}-\frac{1}{(m\omega_1+n\omega_2)^2}\end{equation}
is a doubly-periodic elliptic function of $z\in\mathds{C}$ with periods $\omega_1,\omega_2\in\mathds{C}$ such that $Im(\omega_1/\omega_2)>0$. $\wp$ is the inverse of the elliptic integral
\begin{equation}\wp^{-1}(x; g_2,g_3)=\displaystyle\int_x^\infty \frac{1}{\sqrt{4t^3-g_2t-g_3}} \ dt\end{equation}
where $g_2, g_3\in\mathds{C}$ are known as \emph{Weierstrass invariants}.  
Given periods $\omega_1,\omega_2$, the invariants are
\begin{eqnarray}g_2&=&60\hspace{-.15in}\displaystyle\sum_{(m,n)\in\mathds{Z}\times\mathds{Z}-\{(0,0)\}}\frac{1}{(m\omega_1+n\omega_2)^4}\\
g_3&=&140\displaystyle\sum_{(m,n)\in\mathds{Z}\times\mathds{Z}-\{(0,0)\}}\frac{1}{(m\omega_1+n\omega_2)^6}\notag.\end{eqnarray}
Alternately given invariants $g_2,g_3$, periods $\omega_1,\omega_2$ can be constructed if  the \emph{discriminant} $\Delta\stackrel{def.}{=}g_2^3-27g_3^2$ is nonzero -- that is, when the \emph{Weierstrass cubic} $4t^3-g_2t-g_3$ does not have repeated roots (see $21\cdot{73}$ of Whittaker and Watson \cite{WW}).  In this lecture we refer to $\wp(z;\omega_1,\omega_2)$ as either $\wp(z)$ or $\wp(z;g_2,g_3)$ and consider only cases where the invariants are such that $g_2^3\neq 27g_3^2$.  By (2.2) and (2.7) one can see that the Weierstrass elliptic function  satisfies 
\begin{equation}\wp'(z)^2=4\wp(z)^3-g_2\wp(z)-g_3.\end{equation}
Note that in Michael Tuite's lecture in this volume,  $\omega_{m,n}$ there is equal to $m\omega_1+n\omega_2$ here with our $\omega_1,\omega_2$ specialized to $2\pi i \tau$ and $2\pi i$ in his lecture.

In the special case that the discriminant $\Delta>0$, the roots of the Weierstrass cubic $4t^3-g_2t-g_3$ are real and distinct, and are conventionally notated by $e_1>e_2>e_3$ for $e_1+e_2+e_3=0$ (since the cubic contains no squared term).  In this case $4t^3-g_2t-g_3=4(t-e_1)(t-e_2)(t-e_3)$ so that the Weierstrass invariants are given in terms of the roots by
\begin{equation}g_2=-4(e_1e_2+e_1e_3+e_2e_3),\qquad g_3=4e_1e_2e_3.\end{equation}
For $e_1<z\in\mathds{R}$, $\wp$ can then be written in terms of the Jacobi elliptic functions in any one of the following ways
\begin{eqnarray}\wp(z; g_2, g_3)&=&e_3+\gamma^2 ns^2(\gamma z, k)\notag\\
\wp(z; g_2, g_3)&=&e_2+\gamma^2 ds^2(\gamma z,k)\\
\wp(z; g_2, g_3)&=&e_1+\gamma^2 cs^2(\gamma z, k)\notag\end{eqnarray}
where $\gamma^2\stackrel{def.}{=}e_1-e_3$ and the modulus $k$ is such that $k^2=\frac{e_2-e_3}{e_1-e_3}$ (similar equations hold if $z\in\mathds{R}$ is in a different range in relation to the real roots $e_1, e_2, e_3$, and alternate relations hold for non-real roots when $\Delta<0$,  see chapter II of Greenhill \cite{Greenhill}).  

Note that the Jacobi elliptic functions solve a differential equation which contains only even powers of $f(u)$, and $\wp$ solves an equation with no squared or quartic powers of $\wp$.  Weierstrass elliptic functions have the advantage of being easily implemented in the case that the cubic $4t^3-g_2t-g_3$ is not factored in terms of its roots.  Elliptic integrals of type $\mbox{\large{$\int$}}R(x)/\sqrt{P(x)} dx$, where $P(x)$ is a cubic polynomial and $R$ is a rational function of $x$, can be written in terms of three fundamental \emph{Weierstrassian normal elliptic integrals} although we will not record the details here (see Appendix of Byrd and Friedman \cite{BF}).  In section 6 we will see a method which allows one to write the elliptic integral $\int_{x_0}^x 1/\sqrt{F(t)} dt$, for $F(t)$ a quartic polynomial,  in terms of the Weierstrassian normal elliptic integral \emph{of the first kind} (2.7) by reducing the quartic to a cubic.  

\newpage

\section{\protect Jacobi theta functions}

\indent\indent \emph{Jacobi theta functions} are functions of two arguments, $z\in\mathds{C}$ a complex number and $\tau\in\mathds{H}$ in the upper-half plane.  Every elliptic function can be written as the ratio of two theta functions. Doing so elucidates the meromorphic nature of elliptic functions and is useful in the numerical evaluation of elliptic functions.  One must be cautious with the notation of theta functions, since many different conventions are used.  We will use the notation of Whittaker and Watson \cite{WW} to define
\begin{eqnarray}\theta_1(z,\tau)&\stackrel{def.}{=}&2q^{1/4}\displaystyle\sum_{n=0}^\infty (-1)^nq^{n(n+1)}sin((2n+1)z)\notag\\
\theta_2(z,\tau)&\stackrel{def.}{=}&2q^{1/4}\displaystyle\sum_{n=0}^\infty q^{n(n+1)}cos((2n+1)z)\\
\theta_3(z,\tau)&\stackrel{def.}{=}&1+2\displaystyle\sum_{n=1}^\infty q^{n^2}cos(2nz)\notag\\
\theta_4(z,\tau)&\stackrel{def.}{=}&1+2\displaystyle\sum_{n=1}^\infty (-1)^nq^{n^2}cos(2nz).\notag\end{eqnarray}
where $q\stackrel{def.}{=}e^{\pi i \tau}$ is called the \emph{nome}.  We also define the special values $\theta_i\stackrel{def.}{=}\theta_i(0,\tau)$.  

In terms of theta functions, the Jacobi elliptic functions are 
\begin{eqnarray}sn(u,k)&=&\frac{\theta_3 \ \theta_1(u/\theta_3^2,\tau)}{\theta_2 \  \theta_4(u/\theta_3^2,\tau)}\notag\\
cn(u,k)&=&\frac{\theta_4 \ \theta_2(u/\theta_3^2,\tau)}{\theta_2 \ \theta_4(u/\theta_3^2,\tau)}\\
dn(u,k)&=&\frac{\theta_4 \ \theta_3(u/\theta_3^2,\tau)}{\theta_3 \ \theta_4(u/\theta_3^2,\tau)}\notag\end{eqnarray}
where $\tau$ is chosen such that $k^2=\theta_2^4/\theta_3^4$.  By $22\cdot 11$ of Whittaker and Watson \cite{WW}, if 
$0<k^2<1$, there exists a value of $\tau$ for which the quotient  $\theta_2^4/\theta_3^4=k^2$.

\newpage
\section{\protect The FRLW cosmological model}
\indent\indent The Friedmann-Robertson-Lema\^itre-Walker cosmological model assumes that our current expanding universe is on large scales homogeneous and isotropic.  On a $d+1-$dimensional spacetime  this assumption translates into a metric of the form 
\begin{equation}ds^2=-dt^2+\widetilde{a}(t)^2\left(\frac{dr^2}{1-k'r^2}+r^2d\Omega^2_{d-1}\right)\end{equation}
where $\widetilde{a}(t)$ is the \emph{cosmic scale factor} and $k'\in\{-1,0,1\}$ is the \emph{curvature parameter}.  

Einstein field equations $G_{ij}=-\kappa_d T_{ij}+\Lambda g_{ij}$ then govern the evolution of the universe over time.  In these equations, the Einstein tensor $G_{ij}\stackrel{def.}{=}R_{ij}-\frac{1}{2}Rg_{ij}$ is computed directly from the metric $g_{ij}$ by calculating the Ricci tensor $R_{ij}$ and the scalar curvature $R$.  Also $\kappa_d=8\pi G_d$ where $G_d$ is a generalization of Newton's constant to $d+1-$dimensional spacetime and $\Lambda>0$ is the cosmological constant.  The form of the energy-momentum tensor $T_{ij}$ depends on what sort of matter content one is assuming, and in this lecture will be that of a perfect fluid -- that is, $T_{ij}=(p+\rho)g_{i0}g_{j0}+pg_{ij}$ where $\rho(t)$ and $p(t)$ are the density and pressure of the fluid respectively.  

For the metric (4.1) with a perfect fluid energy-momentum tensor, Einstein's equations are
\begin{equation*}\frac{d(d-1)}{2}\left(H^2+\frac{k'}{\widetilde{a}^2}\right)\stackrel{(i)}{=} \kappa_d \rho(t)+\Lambda\end{equation*}
\begin{equation*}(d-1)\dot{H}+\frac{d(d-1)}{2}H^2+\frac{(d-1)(d-2)}{2}\frac{k'}{\widetilde{a}^2}\stackrel{(ii)}{=}-\kappa_d p(t)+\Lambda.\end{equation*}
for $H(t)\stackrel{def.}{=}{\dot{\widetilde{a}}(t)}/{\widetilde{a}(t)}$ and where dot denotes differentiation with respect to $t$.  In this lecture, only equation (i) will be required to relate the cosmic scale factor $\widetilde{a}(t)$ to elliptic and theta functions.  

We begin by rewriting equation (i) in terms of \emph{conformal time} $\eta$ by defining the \emph{conformal scale factor} $a(\eta)\stackrel{def.}{=}\widetilde{a}(f(\eta))$ where $f(\eta)$ is the inverse function of $\eta(t)$ which satisfies $\dot\eta(t)=1/\widetilde{a}(t)$.  In terms of $a(\eta)$, (i) becomes
\begin{equation}a'(\eta)^2=\widetilde{\Lambda}a(\eta)^4+\widetilde{\kappa}_d \rho(f(\eta))a(\eta)^4-k'a(\eta)^2\end{equation}
where we use notation $\widetilde{\Lambda}\stackrel{def.}{=}2\Lambda/d(d-1)$, $\widetilde{\kappa}_d\stackrel{def.}{=}2\kappa_d/d(d-1)$ and we take spatial dimension $d>1$.

\newpage
\section{\protect FRLW and Jacobi elliptic and theta\\ functions}

\indent\indent  In general, if $f(u)$ is a solution to $f'(u)^2+af(u)^2+bf(u)^4=c$, then $g(u)=\beta f(\alpha u)$ is a solution to 
\begin{equation}g'(u)^2+Ag(u)^2+Bg(u)^4=\frac{A^2bc}{a^2B}\end{equation}
for $\alpha=\sqrt{\frac{A}{a}}$ and $\beta=\sqrt{\frac{Ab}{aB}}$ where we may choose either the positive or negative square root for each of $\alpha$ and $\beta$.  We will construct solutions to the conformal time FRLW Einstein equation (4.2), given that Jacobi elliptic functions solve (2.5), and also proceed to write these solutions in terms of theta functions by the relations in (3.2).   
   
For the special case of density $\rho(t)=\frac{D}{\widetilde{a}(t)^4}$ with $D>0$, (4.2) becomes
\begin{equation}a'(\eta)^2+k' a(\eta)^2-\widetilde{\Lambda}a(\eta)^4=\widetilde{\kappa}_d D.\end{equation}
We construct real-valued solutions to this equation as above in terms of Jacobi elliptic functions, by restricting to special values of $k'$ and $D=\frac{A^2bc}{a^2B\widetilde\kappa_d}>0$ for which the ratios ${a/A}$ and ${b/B}$ are positive. 

In the case of positive curvature $k'=1$ and $D=\frac{k^2}{\widetilde{\kappa_d}\stackrel{}{\widetilde{\Lambda}}(1+k^2)^2}$ for any $0<k<1$, the conformal time equation (5.2) becomes 
\begin{equation}a'(\eta)^2+a(\eta)^2-\widetilde{\Lambda}a(\eta)^4=\frac{k^2}{\stackrel{}{\widetilde{\Lambda}}(1+k^2)^2}\end{equation}
so that  $A=1$ and $B=-\widetilde{\Lambda}$.  By considering Jacobi elliptic functions for which $\frac{A}{a}=\frac{1}{a}$ and $\frac{B}{b}=\frac{-\widetilde\Lambda}{b}$ are positive, some solutions to (5.3) in terms of Jacobi functions (and their equivalent theta function representations) are
\begin{eqnarray}\\
a_{sn}(\eta)\hspace{-2mm}
&=&\hspace{-2mm}
\frac{k}{\sqrt{\widetilde{\Lambda}(1+k^2)}} \ sn\left(\frac{\eta}{\sqrt{1+k^2}},k\right)
=
\frac{\theta_2\theta_3}{\sqrt{\widetilde{\Lambda}(\theta_3^4+\theta_2^4)}} \ \frac{\theta_1(\eta/\sqrt{\theta_3^4+\theta_2^4},\tau)}{\theta_4(\eta/\sqrt{\theta_3^4+\theta_2^4},\tau)}\notag\\
a_{ns}(\eta)\hspace{-2mm}
&=&\hspace{-2mm}
\frac{1}{\sqrt{\widetilde{\Lambda}(1+k^2)}} \ ns\left(\frac{\eta}{\sqrt{1+k^2}},k\right)
=
\frac{\theta_2\theta_3}{\sqrt{\widetilde{\Lambda}(\theta_3^4+\theta_2^4)}} \ \frac{\theta_4(\eta/\sqrt{\theta_3^4+\theta_2^4},\tau)}{\theta_1(\eta/\sqrt{\theta_3^4+\theta_2^4},\tau)}\notag\end{eqnarray}
\begin{eqnarray}
a_{cd}(\eta)\hspace{-2mm}
&=&\hspace{-2mm}
\frac{k}{\sqrt{\widetilde{\Lambda}(1+k^2)}} \ cd\left(\frac{\eta}{\sqrt{1+k^2}},k\right)
=
\frac{\theta_2\theta_3}{\sqrt{\widetilde{\Lambda}(\theta_3^4+\theta_2^4)}} \ \frac{\theta_2(\eta/\sqrt{\theta_3^4+\theta_2^4},\tau)}{\theta_3(\eta/\sqrt{\theta_3^4+\theta_2^4},\tau)}\notag\\
a_{dc}(\eta)\hspace{-2mm}
&=&\hspace{-2mm}
\frac{1}{\sqrt{\widetilde{\Lambda}(1+k^2)}} \ dc\left(\frac{\eta}{\sqrt{1+k^2}},k\right)
=
\frac{\theta_2\theta_3}{\sqrt{\widetilde{\Lambda}(\theta_3^4+\theta_2^4)}} \ \frac{\theta_3(\eta/\sqrt{\theta_3^4+\theta_2^4},\tau)}{\theta_2(\eta/\sqrt{\theta_3^4+\theta_2^4},\tau)}\notag
\end{eqnarray}
where $\tau$ is chosen such that $k^2=\theta_2^4/\theta_3^4$.
 
 The first two solutions, $a_{sn}(\eta)$ and $a_{ns}(\eta)$, reduce to hyperbolic trigonometric functions in the case of modulus $k=1$, since $sn(u,1)=tanh(u)$ and $ns(u,1)=coth(u)$.  That is, two additional solutions in terms of elementary functions are
\begin{eqnarray}a_1(\eta)&=&\frac{1}{\sqrt{2\widetilde{\Lambda}}} \ tanh(\eta/\sqrt{2})\\
a_2(\eta)&=&\frac{1}{\sqrt{2\widetilde{\Lambda}}} \ coth(\eta/\sqrt{2}).\notag\end{eqnarray}
For these two solutions, one may solve the differential equation $\dot\eta(t)=1/\widetilde{a}(t)=1/a(\eta(t))$ for $\eta(t)$, and therefore obtain the cosmic scale factor $\widetilde{a}(t)=a(\eta(t))$ which solves the Einstein field equation (i) in section 4 for $\rho(t)=D/\widetilde{a}(t)^4$ with special values $k'=1$ and $D=k^2/\widetilde\Lambda\widetilde\kappa_d(1+k^2)^2$.  Doing so, we obtain 
\begin{eqnarray}\\
\notag\widetilde{a}_1(t)=a_1(\eta(t))
&=&
\frac{1}{\sqrt{2\widetilde{\Lambda}}}  \ tanh\left(ln\left(e^{\sqrt{\widetilde{\Lambda}}t}+\sqrt{e^{2\sqrt{\widetilde{\Lambda}}t}-1} \ \right)\right)\\
\widetilde{a}_2(t)=a_2(\eta(t))
&=&
\frac{1}{\sqrt{2\widetilde{\Lambda}}} \ coth\left(ln\left(e^{\sqrt{\widetilde{\Lambda}}t}+\sqrt{e^{2\sqrt{\widetilde{\Lambda}}t}+1} \ \right)\right)\notag\end{eqnarray}
for $t>0$.

In the case of negative curvature $k'=-1$ and $D=\frac{1-k^2}{\stackrel{}{\widetilde{\Lambda}}\widetilde{\kappa}_d(k^2-2)^2}$ for any $0<k<1$, the conformal time equation (5.2) becomes 
\begin{equation}a'(\eta)^2-a(\eta)^2-\widetilde{\Lambda}a(\eta)^4=\frac{1-k^2}{(k^2-2)^2\widetilde{\Lambda}}.\end{equation}
so that $A=-1$ and $B=-\widetilde\Lambda$.  Again considering Jacobi elliptic functions for which $a,b$ are such that $\frac{A}{a}=\frac{-1}{a}$ and $\frac{B}{b}=\frac{-\widetilde\Lambda}{b}$ are positive and also $D=\frac{A^2bc}{a^2B\widetilde\kappa_d}$, two solutions to (5.7) in terms of Jacobi elliptic functions (and their equivalent theta function representations) are
\begin{eqnarray}
a_{sc}(\eta)
&=&
\sqrt{\frac{1-k^2}{\widetilde{\Lambda}(2-k^2)}} \ sc\left(\frac{\eta}{\sqrt{2-k^2}},k\right)\\
\notag\\
&=&
\sqrt{\frac{\theta_3^4-\theta_2^4}{\widetilde{\Lambda}(2\theta_3^4-\theta_2^4)}} \ \frac{\theta_3 \ \theta_1(\eta/\sqrt{2\theta_3^4-\theta_2^4},\tau)}{\theta_4 \ \theta_2(\eta/\sqrt{2\theta_3^4-\theta_2^4},\tau)}\notag\\
\notag\\
a_{cs}(\eta)
&=&
\frac{1}{\sqrt{\widetilde{\Lambda}(2-k^2)}} \ cs\left(\frac{\eta}{\sqrt{2-k^2}},k\right)\notag\\
\notag\\
&=&
\frac{\theta_3\theta_4}{\sqrt{\widetilde{\Lambda}(2\theta_3^4-\theta_2^4)}} \ \frac{ \theta_2(\eta/\sqrt{2\theta_3^4-\theta_2^4},\tau)}{ \theta_1(\eta/\sqrt{2\theta_3^4-\theta_2^4},\tau)}
\notag\end{eqnarray}
where $\tau$ is such that $k^2=\theta_2^4/\theta_3^4$.  

 Note that by the comments following equation (2.11), it is possible to express the solutions obtained in this section in terms of Weierstrass functions.  E. Abdalla and L. Correa-Borbonet \cite{ACB} have also considered the Einstein equation (i) with $\rho(t)=D/\widetilde{a}(t)^4$, and have found connections with Weierstrass functions in cosmic time (as opposed to the conformal time argument given here).  In section 6 of this lecture we will find more general solutions to the conformal time equation (4.2) in terms of $\wp$, for arbitrary curvature $k'$ and positive $D$-value.  We will also consider the additional density functions $\rho(t)=D/\widetilde{a}(t)^3$ and $\rho(t)=D_1/\widetilde{a}(t)^3+D_2/\widetilde{a}(t)^4$ for $D,D_1,D_2>0$.  

\newpage
\section{\protect FRLW and Weierstrass elliptic functions}

\indent\indent In general, if $g(0)=x_0$ and $g(u)$ satisfies
\begin{equation}g'(u)^2=F(g(u))\end{equation}
for $F(x)=A_4x^4+A_3x^3+A_2x^2+A_1x+A_0$ any quartic polynomial with no repeated roots, then the inverse function $y(x)$ of $g(u)$ is the elliptic integral
\begin{equation}y(x)=\displaystyle\int_{x_0}^x \frac{dt}{\sqrt{F(t)}}.\end{equation}

{\bf For initial condition \boldmath{$g'(0)=0$}}, by (6.1) $x_0$ is a root of the polynomial $F(x)$.  In this case the integral (6.2) can be rewritten as $\int_\xi^\infty dz/\sqrt{P(z)}$ where $\xi=1/(x-x_0)$ and $P(z)$ is a cubic polynomial.  To do this, first expand $F(t)$ into its Taylor series about $x_0$ and then perform the change of variables $z=1/(t-x_0)$.  Furthermore one may obtain the form $\int_\tau^\infty d\sigma/\sqrt{Q(\sigma)}$ for $\tau=\frac{1}{4}(F'(x_0)\xi +\frac{1}{6} F''(x_0))$ where $Q(\sigma)=4\sigma^3-g_2\sigma-g_3$ is a Weierstrass cubic.  This is done  by setting $z=(4\sigma-B_2/3)/B_3$ where $B_2=F''(x_0)/2$ and $B_3=F'(x_0)$ are the quadratic and cubic coefficients of $P(z)$ respectively.  Note that since $F(x)$ has no repeated roots, $x_0$ is not a double root, $F'(x_0)\neq 0$ and the variable $z$ is well-defined.  Therefore we have obtained $y\left(\frac{6F'(x_0)}{24\tau-F''(x_0)}+x_0\right)=\wp^{-1}(\tau)$.  Writing this in terms of $x$, setting $x=g(u)$ and solving for $g(u)$, one obtains the solution to (6.1)
\begin{equation}g(u)=x_0+\frac{F'(x_0)}{4\wp(u; g_2, g_3)-\frac{F''(x_0)}{6}}\end{equation}
where 
\begin{eqnarray}g_2&=&A_0A_4-\frac{1}{4}A_1A_3+\frac{1}{12}A_2^2\\
g_3&=&\frac{1}{6}A_0A_2A_4+\frac{1}{48}A_1A_2A_3-\frac{1}{16}A_1^2A_4-\frac{1}{16}A_0A_3^2-\frac{1}{216}A_2^3\notag\end{eqnarray}
are referred to as the \emph{invariants of the quartic} $F(x)$.  Since $F(x)$ has no repeated roots, the discriminant $\Delta=g_2^3-27g_3^2\neq 0$.  Here if $x_0=0$, (6.3) becomes
\begin{equation}g(u)=\frac{A_1}{4\wp(u;g_2,g_3)-\frac{A_2}{3}}.\end{equation}

{\bf For initial condition \boldmath{$ g'(0)\neq 0$}}, by (6.1) $x_0$ is not a root of $F(x)$ and a more general solution to (6.1) is due to Weierstrass.  The proof (which we will not include here) was published by Biermann in 1865 (see \cite{Biermann, Reynolds}).  The solution is\\
 \begin{equation}g(u)=x_0+\frac{\sqrt{F(x_0)}\wp'(u)+\frac{1}{2}F'(x_0)\left(\wp(u)-\frac{1}{24}F''(x_0)\right)+\frac{1}{24}F(x_0)F'''(x_0)}{2\left(\wp(u)-\frac{1}{24}F''(x_0)\right)^2-\frac{1}{48}F(x_0)F^{(iv)}(x_0)}\end{equation}
 where $\wp$ is formed with the invariants of the quartic seen in (6.4) such that $\Delta\neq 0$.  Here if $x_0=0$, (6.6) becomes
 \begin{equation}g(u)=\frac{\sqrt{A_0}\wp'(u)+\frac{1}{2}A_1\left(\wp(u)-\frac{1}{12}A_2\right)+\frac{1}{4}A_0A_3}{2\left(\wp(u)-\frac{1}{12}A_2\right)^2-\frac{1}{2}A_0A_4}.\end{equation}

As a first example we consider the conformal time Einstein equation (4.2) for density $\rho(t)=D_1/\widetilde{a}(t)^3+D_2/\widetilde{a}(t)^4$ with $D_1,D_2>0$.  In this case (4.2) becomes
\begin{equation}a'(\eta)^2 =\widetilde{\Lambda}a(\eta)^4-k' a(\eta)^2+ \widetilde{\kappa}_d D_1 a(\eta)+ \widetilde{\kappa}_d D_2 \end{equation}
and we take $A_4=\widetilde{\Lambda}, A_3=0, A_2=-k',  A_1=\widetilde{\kappa}_d D_1$ and $A_0=\widetilde{\kappa}_d D_2$.  

The most general solution to (6.8) seen here is with initial conditions $a'(0)\neq 0$ so that $a(0)\stackrel{def.}{=}a_0$ is not a root of the polynomial
\begin{equation}F(t)=\widetilde{\Lambda}t^4-k't^2+\widetilde{\kappa}_d D_1 t +\widetilde{\kappa}_d D_2.\end{equation}
In this case the solution to (6.8) is given by (6.6) as
 \begin{equation}a(\eta)=a_0+\frac{\sqrt{F(a_0)}\wp'(\eta)+\frac{1}{2}F'(a_0)\left(\wp(\eta)-\frac{1}{24}F''(a_0)\right)+\frac{1}{24}F(a_0)F'''(a_0)}{2\left(\wp(\eta)-\frac{1}{24}F''(a_0)\right)^2-\frac{1}{48}F(a_0)F^{(iv)}(a_0)}\end{equation}
with Weierstrass invariants
\begin{eqnarray}g_2&=&\widetilde{\Lambda}\widetilde{\kappa}_d D_2+\frac{(k')^2}{12}\\
g_3&=&-\frac{1}{6}\widetilde{\Lambda}\widetilde{\kappa}_d D_2   k' -\frac{1}{16}\widetilde{\Lambda}\widetilde{\kappa}_d^2 D_1^2+\frac{(k')^3}{216}\notag\end{eqnarray}
restricted to be such that $\Delta=g_2^3-27g_3^2\neq 0$ so that $F(t)$ does not have repeated roots.  
\newpage
Since $D_2,\widetilde\kappa_d>0$, by (6.9) zero is not a root of $F(t)$ and therefore for initial conditions $a'(0)\neq 0$ and $a_0=0$ the solution to (6.8) is given by (6.7),
 \begin{equation}a(\eta)=\frac{\sqrt{\widetilde{\kappa}_d D_2}\wp'(\eta)+\frac{1}{2}\widetilde\kappa_d D_1\left(\wp(\eta)+\frac{k'}{12}\right)}{2\left(\wp(\eta)+\frac{k'}{12}\right)^2-\frac{1}{2}\widetilde{\Lambda} \widetilde{\kappa}_d  D_2}\end{equation}
for  invariants $g_2,g_3$ as in (6.11) with $\Delta\neq 0$.  One can compare this with the results in papers by Aurich, Steiner and Then, where curvature is taken to be $k'=-1$ \cite{AurichSteiner, AurichSteinerThen}.

Finally, for $a'(0)=0$ and $a_0\neq 0$ a root of $F(t)$ in (6.9), the solution to (6.8) is given by (6.3),
\begin{equation}a(\eta)=a_0+\frac{F'(a_0)}{4\wp(\eta)-\frac{F''(a_0)}{6}}\end{equation}
again with invariants (6.11) such that $\Delta\neq 0$.

For a more concrete example, consider the density function $\rho(t)=D/\widetilde{a}(t)^3$ for $D>0$ so that conformal time equation (4.2) becomes
\begin{equation}a'(\eta)^2=\widetilde{\Lambda}a(\eta)^4-k'a(\eta)^2+\widetilde{\kappa}_d Da(\eta).\end{equation}
Here zero is a root of the polynomial $F(t)$ with $A_4=\widetilde\Lambda, A_3=A_0=0, A_2=-k'$ and $A_1=\widetilde{\kappa}_d D$.  Therefore with initial conditions $a'(0)=a(0)=0$, the solution to (6.14) is given by (6.5) as 
\begin{equation}a(\eta)=\frac{3\widetilde\kappa_d D}{12\wp(\eta)+k'}\end{equation}
with invariants
\begin{equation}\quad g_2=\frac{(k')^2}{12}\qquad\mbox{ and }\qquad
g_3=-\frac{1}{16}\widetilde{\Lambda}\widetilde{\kappa}_d^2D^2+\frac{(k')^3}{216}\end{equation}
restricted to be such that $\Delta\neq 0$.  

As noted in the comments following equations (2.11), one can write this solution in terms of Jacobi elliptic functions (by using equations (2.11) if the roots of the reduced cubic $4t^3-g_2t-g_3$ are real).  To demonstrate this, we choose $D=\frac{1}{3\widetilde\kappa_d}\sqrt{\frac{2}{3\widetilde\Lambda}}$ and $k'=1$ so that  $g_3=0$ and $g_2=1/12$.  For this positive curvature case,  (6.15) becomes
\begin{equation}a(\eta)=\frac{\sqrt{2}}{\sqrt{3\widetilde\Lambda}\left(12\wp(\eta)+1\right)}\end{equation}
and the reduced cubic is $4t^3-(1/12)t=4t(t-1/4\sqrt{3})(t+1/4\sqrt{3})$.  Applying (2.11) with $e_3=-1/4\sqrt{3}, e_2=0, e_1=1/4\sqrt{3}$, (6.17) can be equivalently written in terms of Jacobi elliptic functions for $\eta>1/4\sqrt{3}$ as 
\begin{eqnarray}a(\eta)&=&\frac{\sqrt{2/\widetilde\Lambda}}{\sqrt{3}-3+6ns^2\left(\mbox{\large{$\frac{\eta}{\sqrt{2\sqrt{3}}}$}},\mbox{\large{$\frac{1}{\sqrt{2}}$}}\right)}\\
&=&\frac{\sqrt{2/\widetilde\Lambda}}{\sqrt{3}+6ds^2\left(\mbox{\large{$\frac{\eta}{\sqrt{2\sqrt{3}}}$}}, \mbox{\large{$\frac{1}{\sqrt{2}}$}}\right)}\notag\\
&=&\frac{\sqrt{2/\widetilde\Lambda}}{\sqrt{3}+3+6cs^2\left(\mbox{\large{$\frac{\eta}{\sqrt{2\sqrt{3}}}$}},\mbox{\large{$\frac{1}{\sqrt{2}}$}}\right)}\notag
\notag   
\end{eqnarray}
since $k^2=\frac{e_2-e_3}{e_1-e_3}=\frac{1}{2}$ and  $\gamma^2=e_1-e_3=1/2\sqrt{3}$.  In terms of theta functions, (6.18) becomes
\begin{eqnarray}\\ \notag a(\eta)&=&\frac{\sqrt{2/\widetilde\Lambda} \ {\theta_3^2 \  \theta_1^2(\eta/\sqrt{2\sqrt{3}} \ \theta_3^2, \tau)}}{(\sqrt{3}-3) \ {\theta_3^2 \  \theta_1^2(\eta/\sqrt{2\sqrt{3}} \ \theta_3^2, \tau)}+6 \ {\theta_2^2 \ \theta_4^2(\eta/\sqrt{2\sqrt{3}} \ \theta_3^2, \tau)}}\\
\notag\\
&=&\frac{\sqrt{2/\widetilde\Lambda} \ \theta_3^4 \  \theta_1^2(\eta/\sqrt{2\sqrt{3}} \ \theta_3^2, \tau) }{\sqrt{3} \ \theta_3^4 \  \theta_1^2(\eta/\sqrt{2\sqrt{3}} \ \theta_3^2, \tau)+6 \ \theta_2^2  \ \theta_4^2 \  \theta_3^2(\eta/\sqrt{2\sqrt{3}} \ \theta_3^2, \tau)}\notag\\
\notag\\
&=&\frac{\sqrt{2/\widetilde\Lambda} \ \theta_3^2 \ \theta_1^2(\eta/\sqrt{2\sqrt{3}} \ \theta_3^2, \tau)}{(\sqrt{3}+3) \ \theta_3^2 \ \theta_1^2(\eta/\sqrt{2\sqrt{3}} \ \theta_3^2, \tau)+6 \ \theta_4^2 \ \theta_2^2(\eta/\sqrt{2\sqrt{3}} \ \theta_3^2, \tau)}\notag
\end{eqnarray}
where  $\tau$ is taken such that $1/2=\theta_2^4/\theta_3^4$.

As a final example, we return to $\rho(t)=D/\widetilde{a}(t)^4$ for $D>0$, which was considered in section 5.  That is, we will obtain alternate solutions to equation (5.2).  Since $D>0$, zero is not a root of the polynomial $F(t)$ with $A_4=\widetilde\Lambda, A_3=A_1=0, A_2=-k'$ and $A_0=\widetilde\kappa_d D$.  Therefore for initial conditions $a'(0)\neq 0$ and $a(0)=0$, (6.7) gives the solution
\begin{equation}a(\eta)=\frac{\sqrt{\widetilde\kappa_d D}\wp'(\eta)}{2(\wp(\eta)+\frac{k'}{12})^2-\frac{1}{2}\widetilde\Lambda \widetilde\kappa_d D}\end{equation}
for invariants 
\begin{equation}g_2=\widetilde\Lambda \widetilde\kappa_d D +\frac{(k')^2}{12}\qquad\mbox{ and }\qquad g_3=-\frac{1}{6}\widetilde\Lambda\widetilde\kappa_d D k' + \frac{(k')^3}{216}\end{equation}
restricted to be such that $\Delta\neq 0$.  (6.20) is more general than the solutions to (5.2) in section 5, since here the curvature $k'$ and the constant $D$ are unspecified.    

To see this solution expressed in terms of Jacobi elliptic functions, take curvature $k'=1$ and $D=\frac{1}{36\widetilde\Lambda \widetilde\kappa_d}$ so that $g_3=0$ and $g_2=1/9$.   Then the reduced cubic is $4t^3-(1/9)t=4(t-1/6)(t+1/6)$ so that $e_3=-1/6, e_2=0, e_1=1/6$ and (6.20) becomes
\begin{equation}a(\eta)=\frac{\frac{12}{\sqrt{\widetilde\Lambda}} \  \wp ' (\eta)}{144 (\wp(\eta)+\frac{1}{12})^2-1}.\end{equation}
Using (2.9) and (2.11) to write (6.22) in terms of Jacobi elliptic functions for $\eta>1/6$,
\begin{eqnarray}\\
\notag a(\eta)&=&\frac{\sqrt{2-3sn^2\left(\frac{\eta}{\sqrt{3}},\frac{1}{\sqrt{2}}\right)+sn^4\left(\frac{\eta}{\sqrt{3}},\frac{1}{\sqrt{2}}\right)}}{\sqrt{6\widetilde\Lambda}\left(2ns\left(\frac{\eta}{\sqrt{3}},\frac{1}{\sqrt{2}}\right)-sn\left(\frac{\eta}{\sqrt{3}},\frac{1}{\sqrt{2}}\right)\right)}\\
\notag\\
&=&\frac{1}{2\sqrt{3\widetilde\Lambda}ds\left(\frac{\eta}{\sqrt{3}},\frac{1}{\sqrt{2}}\right)}\sqrt{\frac{2ds^2\left(\frac{\eta}{\sqrt{3}},\frac{1}{\sqrt{2}}\right)-1}{2ds^2\left(\frac{\eta}{\sqrt{3}},\frac{1}{\sqrt{2}}\right)+1}}\notag\\
\notag\\
&=&\frac{cs\left(\frac{\eta}{\sqrt{3}},\frac{1}{\sqrt{2}}\right)}
{\sqrt{6\widetilde\Lambda}\sqrt{1+3cs^2\left(\frac{\eta}{\sqrt{3}},\frac{1}{\sqrt{2}}\right)+2cs^4\left(\frac{\eta}{\sqrt{3}},\frac{1}{\sqrt{2}}\right)}}\notag
\end{eqnarray}
where each of the positive and negative square roots solve (5.2) for $k'=1, D=1/36\widetilde\Lambda\widetilde\kappa_d$ and where $\gamma^2=e_1-e_3=1/3$ and modulus $k^2=\frac{e_2-e_3}{e_1-e_3}=1/2$.  \\
\newpage

\noindent Writing (6.23) in terms of theta functions, $a(\eta)$ is 
\begin{equation}\end{equation}
\begin{eqnarray}
=\frac{\theta_3 \theta_1(\eta/\sqrt{3}\theta_3^2)\sqrt{2\theta_2^4\theta_4^4(\eta/\sqrt{3}\theta_3^2)-3\theta_2^2\theta_3^2\theta_1^2(\eta/\sqrt{3}\theta_3^2)\theta_4^2(\eta/\sqrt{3}\theta_3^2)+\theta_3^4\theta_1^4(\eta/\sqrt{3}\theta_3^2)}}{\sqrt{6\widetilde\Lambda} \ \theta_2\theta_4(\eta/\sqrt{3}\theta_3^2)\left(2\theta_2^2\theta_4^2(\eta/\sqrt{3}\theta_3^2)-\theta_3^2\theta_1^2(\eta/\sqrt{3}\theta_3^2)\right)}\notag\end{eqnarray}
\\
\begin{equation}\notag =\frac{\theta_3^2 \theta_1(\eta/\sqrt{3}\theta_3^2)}{2\sqrt{3\widetilde\Lambda} \ \theta_2 \theta_4 \theta_3(\eta/\sqrt{3}\theta_3^2)}\sqrt{\frac{2\theta_2^2\theta_4^2\theta_3^2(\eta/\sqrt{3}\theta_3^2)-\theta_3^4\theta_1^2(\eta/\sqrt{3}\theta_3^2)}{2\theta_2^2\theta_4^2\theta_3^2(\eta/\sqrt{3}\theta_3^2)+\theta_3^4\theta_1^2(\eta/\sqrt{3}\theta_3^2)}}\end{equation}
\\
\begin{equation}\notag 
=\frac{
\theta_4\theta_3\theta_2(\eta/\sqrt{3}\theta_3^2)\theta_1(\eta/\sqrt{3}\theta_3^2)
}{
\sqrt{6\widetilde\Lambda}
\sqrt{
\theta_3^4\theta_1^4(\eta/\sqrt{3}\theta_3^2)
+3\theta_3^2\theta_4^2\theta_1^2(\eta/\sqrt{3}\theta_3^2)\theta_2^2(\eta/\sqrt{3}\theta_3^2)
+2\theta_4^4\theta_2^4(\eta/\sqrt{3}\theta_3^2)
}
}
\end{equation}
by the theta function representations for each of $sn, ds$ and $cs$ respectively.  Here the dependence on $\tau$ in the theta functions is suppressed and $\tau$ is taken to satisfy $1/2=\theta_2^4/\theta_3^4$.

A similar procedure can be done with (6.20) when $k'=-1$ and $D=\frac{1}{36\widetilde\Lambda \widetilde\kappa_d}$ so that again $g_3=0, g_2=1/9 \Rightarrow e_3=-1/6, e_2=0, e_1=1/6$.  In this case (6.20) is
\begin{equation}a(\eta)=\frac{\frac{12}{\sqrt{\widetilde\Lambda}}\wp '(\eta)}{144(\wp(\eta)-\frac{1}{12})^2-1}.\end{equation}
In terms of Jacobi elliptic functions for $\eta>1/6$,
\begin{eqnarray}&&\\
a(\eta)&=&\frac{
sn\left(\frac{\eta}{\sqrt{3}}, \frac{1}{\sqrt{2}}\right)
}
{
\sqrt{6\widetilde\Lambda}\sqrt{sn^4\left(\frac{\eta}{\sqrt{3}}, \frac{1}{\sqrt{2}}\right)-3sn^2\left(\frac{\eta}{\sqrt{3}}, \frac{1}{\sqrt{2}}\right)+2}
}  \notag\\
&=& 
\frac{1}{2\sqrt{3\widetilde\Lambda } ds\left(\frac{\eta}{\sqrt{3}}, \frac{1}{\sqrt{2}}\right) } \sqrt{\frac{2ds^2\left(\frac{\eta}{\sqrt{3}}, \frac{1}{\sqrt{2}}\right)+1}{2ds^2\left(\frac{\eta}{\sqrt{3}}, \frac{1}{\sqrt{2}}\right)-1}}
\notag\\
&=&
\frac{\sqrt{sc^4\left(\frac{\eta}{\sqrt{3}}, \frac{1}{\sqrt{2}}\right)+3sc^2\left(\frac{\eta}{\sqrt{3}}, \frac{1}{\sqrt{2}}\right)+2}}{\sqrt{6\widetilde\Lambda}\left(2cs\left(\frac{\eta}{\sqrt{3}}, \frac{1}{\sqrt{2}}\right)+sc\left(\frac{\eta}{\sqrt{3}}, \frac{1}{\sqrt{2}}\right)\right)}
 \notag\end{eqnarray}
since again $\gamma^2=e_1-e_3=1/3$ and $k^2=\frac{e_2-e_3}{e_1-e_3}=1/2$.  Or, equivalently in terms of theta functions,
\begin{eqnarray}&&\\
a(\eta)&=& 
\frac{
\theta_2\theta_3\theta_1(\eta/\sqrt{3}\theta_3^2)\theta_4(\eta/\sqrt{3}\theta_3^2)
}
{
\sqrt{6\widetilde\Lambda}
\sqrt{
\theta_3^4\theta_1^4(\eta/\sqrt{3}\theta_3^2)-3\theta_2^2\theta_3^2\theta_1^2(\eta/\sqrt{3}\theta_3^2)\theta_4^2(\eta/\sqrt{3}\theta_3^2)+2\theta_2^4\theta_4^4(\eta/\sqrt{3}\theta_3^2)
}
}\notag\\
&&\notag\\
&&\notag\\
&=&
\frac{\theta_3^2 \theta_1(\eta/\sqrt{3}\theta_3^2)}{2\sqrt{3\widetilde\Lambda} \ \theta_2 \theta_4 \theta_3(\eta/\sqrt{3}\theta_3^2)}\sqrt{\frac{2\theta_2^2\theta_4^2\theta_3^2(\eta/\sqrt{3}\theta_3^2)+\theta_3^4\theta_1^2(\eta/\sqrt{3}\theta_3^2)}{2\theta_2^2\theta_4^2\theta_3^2(\eta/\sqrt{3}\theta_3^2)-\theta_3^4\theta_1^2(\eta/\sqrt{3}\theta_3^2)}}
\notag\\
&&\notag\\
&&\notag\\
&=&\sqrt{\theta_3^4\theta_1^4(\eta/\sqrt{3}\theta_3^2)+3\theta_3^2\theta_4^2\theta_1^2(\eta/\sqrt{3}\theta_3^2)\theta_2^2(\eta/\sqrt{3}\theta_3^2)+2\theta_4^4\theta_2^4(\eta/\sqrt{3}\theta_3^2)}\notag\\
&&\notag\\
&&\cdot\frac{\theta_3\theta_1(\eta/\sqrt{3}\theta_3^2)}{\sqrt{6\widetilde\Lambda} \ \theta_4\theta_2(\eta/\sqrt{3}\theta_3^2)\left(2\theta_4^2\theta_2^2(\eta/\sqrt{3}\theta_3^2)+\theta_3^2\theta_1^2(\eta/\sqrt{3}\theta_3^2)\right)}
\notag\end{eqnarray}
where the suppressed $\tau$ is taken to satisfy $1/2=\theta_2^4/\theta_3^4$.
 
 

In (6.20), one could also take $D=\frac{1}{\widetilde\Lambda\widetilde\kappa_d}$ and the invariants (6.21) become $g_2=1+\frac{(k')^2}{12}$ and $g_3=-\frac{1}{6}k'+\frac{(k')^3}{216}$.  If $k'=1$, the roots of the reduced cubic $4t^3-(13/12)t+(35/216)$ are $e_3=-7/12, e_2=1/6, e_1=5/12$ and (6.20) becomes
\begin{equation}a(\eta)=\frac{\wp '(\eta)}{\sqrt{\widetilde\Lambda}\left(2(\wp(\eta)+\frac{1}{12})^2-\frac{1}{2}\right)},\end{equation}
which solves (5.2) in this case.  Writing this solution in terms of Jacobi elliptic functions for $\eta>5/12$,
\begin{eqnarray}
&&\\
a(\eta)&=&\frac{ \sqrt{3sn^4\left( \eta,\frac{\sqrt{3}}{2}\right)-7sn^2\left( \eta,\frac{\sqrt{3}}{2}\right)+4}}{2\sqrt{\widetilde\Lambda}\left(ns\left( \eta,\frac{\sqrt{3}}{2}\right)-sn\left( \eta,\frac{\sqrt{3}}{2}\right)\right)} \notag\end{eqnarray}
\begin{eqnarray}
\qquad\quad&=& \frac{ 2sd\left( \eta,\frac{\sqrt{3}}{2}\right) }{\sqrt{\widetilde\Lambda}\sqrt{-\frac{3}{4}sd^4\left( \eta,\frac{\sqrt{3}}{2}\right)+2sd^2\left( \eta,\frac{\sqrt{3}}{2}\right)+4}}  \notag\\
\notag\\
\qquad\qquad\quad&=&\frac{\sqrt{sc^4\left( \eta,\frac{\sqrt{3}}{2}\right)+5sc^2\left( \eta,\frac{\sqrt{3}}{2}\right)+4}}{2\sqrt{\widetilde\Lambda}\left(sc\left( \eta,\frac{\sqrt{3}}{2}\right)+cs\left( \eta,\frac{\sqrt{3}}{2}\right)\right)}
\notag
\end{eqnarray}
since $\gamma^2=e_1-e_3= 1$ and $k^2=\frac{e_2-e_3}{e_1-e_3}= 3/4$.  Writing these solutions equivalently in terms of theta functions,
\begin{eqnarray}
&&\\
a(\eta)&=&\frac{\theta_3\theta_1(\eta/\theta_3^2)\sqrt{3\theta_3^4\theta_1^4(\eta/\theta_3^2)-7\theta_2^2\theta_3^2\theta_1^2(\eta/\theta_3^2)\theta_4^2(\eta/\theta_3^2)+4\theta_2^4\theta_4^4(\eta/\theta_3^2)}}{2\sqrt{\widetilde\Lambda} \ \theta_2\theta_4(\eta/\theta_3^2) \ \left(\theta_2^2\theta_4^2(\eta/\theta_3^2)-\theta_3^2\theta_1^2(\eta/\theta_3^2)\right) }  \notag\\
&&\notag\\
&=& \frac{
2\theta_2\theta_4\theta_3^2\theta_1(\eta/\theta_3^2)\theta_3(\eta/\theta_3^2)
}{
\sqrt{\widetilde\Lambda}
\sqrt{
-\frac{3}{4}\theta_3^8\theta_1^4(\eta/\theta_3^2)+2\theta_2^2\theta_3^4\theta_4^2\theta_1^2(\eta/\theta_3^2)\theta_3^2(\eta/\theta_3^2)+4\theta_2^4\theta_4^4\theta_3^4(\eta/\theta_3^2)
}
}
 \notag\\
&&\notag\\
&=&\frac{\theta_3\theta_1(\eta/\theta_3^2)\sqrt{\theta_3^4\theta_1^4(\eta/\theta_3^2)+5\theta_3^2\theta_4^2\theta_1^2(\eta/\theta_3^2)\theta_2^2(\eta/\theta_3^2)+4\theta_4^4\theta_2^4(\eta/\theta_3^2)} }{2\sqrt{\widetilde\Lambda} \ \theta_4\theta_2(\eta/\theta_3^2) \ \left(\theta_3^2\theta_1^2(\eta/\theta_3^2)+\theta_4^2\theta_2^2(\eta/\theta_3^2)\right)}\notag
\end{eqnarray}
where the suppressed $\tau$ satisfies $3/4=\theta_2^4/\theta_3^4$.

For $D=\frac{1}{\widetilde\Lambda\widetilde\kappa_d}$ and $k'=0$ the invariants are $g_2=1$ and $g_3=0$ and the roots of the cubic $4t^3-t$ are $e_3=-1/2, e_2=0, e_1=1/2$.  In this case (6.20) becomes
\begin{equation}a(\eta)=\frac{\wp'(\eta)}{\sqrt{\widetilde\Lambda}\left(2\wp(\eta)^2-\frac{1}{2}\right)}\end{equation}
which solves (5.2) in this case.  Writing this solution in terms of Jacobi elliptic functions for $\eta>1/2$,
\begin{eqnarray}
&&\\
a(\eta)&=& \frac{
\sqrt{
sn^4\left(\eta,\frac{1}{\sqrt{2}}\right) -3sn^2\left(\eta,\frac{1}{\sqrt{2}}\right)+2    
}
}{
\sqrt{2\widetilde\Lambda} \left(ns\left(\eta,\frac{1}{\sqrt{2}}\right)-sn\left(\eta,\frac{1}{\sqrt{2}}\right)\right)
}  \notag\end{eqnarray}
\begin{eqnarray}
&=& \frac{2ds\left(\eta,\frac{1}{\sqrt{2}}\right)}{\sqrt{\widetilde\Lambda}\sqrt{4ds^4\left(\eta,\frac{1}{\sqrt{2}}\right)-1}}
  \notag\\
&&\notag\\
\qquad&=&\frac{\sqrt{sc^4\left(\eta,\frac{1}{\sqrt{2}}\right)+3sc^2\left(\eta,\frac{1}{\sqrt{2}}\right)+2}}{\sqrt{2\widetilde\Lambda}\left(cs\left(\eta,\frac{1}{\sqrt{2}}\right)+sc\left(\eta,\frac{1}{\sqrt{2}}\right)\right)}\notag
\end{eqnarray}
since $\gamma^2=e_1-e_3=1$ and $k^2=\frac{e_2-e_3}{e_1-e_3}=1/2$.  In terms of theta functions,
\begin{eqnarray}
&&\\
a(\eta)&=&\frac{\theta_3\theta_1(\eta/\theta_3^2)\sqrt{\theta_3^4\theta_1^4(\eta/\theta_3^2)-3\theta_2^2\theta_3^2\theta_1^2(\eta/\theta_3^2)\theta_4^2(\eta/\theta_3^2)+2\theta_2^4\theta_4^4(\eta/\theta_3^2)}}{\sqrt{2\widetilde\Lambda} \ \theta_2\theta_4(\eta/\theta_3^2)  \left(\theta_2^2\theta_4^2(\eta/\theta_3^2)-\theta_3^2\theta_1^2(\eta/\theta_3^2)\right)}   \notag\\
&&\notag\\
&=& \frac{2\theta_2\theta_4\theta_3^2\theta_1(\eta/\theta_3^2)\theta_3(\eta/\theta_3^2)}{\sqrt{\widetilde\Lambda}\sqrt{4\theta_2^4\theta_4^4\theta_3^4(\eta/\theta_3^2)-\theta_3^8\theta_1^4(\eta/\theta_3^2)}}
 \notag\\
&&\notag\\
&=&\frac{\theta_3\theta_1(\eta/\theta_3^2)\sqrt{\theta_3^4\theta_1^4(\eta/\theta_3^2)+3\theta_3^2\theta_4^2\theta_1^2(\eta/\theta_3^2)\theta_2^2(\eta/\theta_3^2)+2\theta_4^4\theta_2^4(\eta/\theta_3^2)}}{\sqrt{2\widetilde\Lambda} \ \left(\theta_4^2\theta_2^2(\eta/\theta_3^2)+\theta_3^2\theta_1^2(\eta/\theta_3^2)\right)}
\notag
\end{eqnarray}
where $\tau$ is taken to satisfy $1/2=\theta_2^4/\theta_3^4$.

For $D=\frac{1}{\widetilde\Lambda\widetilde\kappa_d}$ and $k'=-1$ the invariants are $g_2=13/12$ and $g_3=35/216$ and the roots of the cubic $4t^3-(13/12)t-(35/216)$ are $e_3=-5/12, e_2=-1/6, e_1=7/12$.  In this case (6.20) is
\begin{equation}a(\eta)=\frac{\wp '(\eta)}{\sqrt{\widetilde\Lambda}\left(2(\wp(\eta)-\frac{1}{12})^2-\frac{1}{2}\right)}\end{equation}
which solves (5.2) in this case. Writing this solution in terms of Jacobi elliptic functions for $\eta>7/12$,
\begin{eqnarray}
&&\\
a(\eta)&=& \frac{\sqrt{sn^4\left(\eta,\frac{1}{2}\right)-5sn^2\left(\eta,\frac{1}{2}\right)+4}}{2\sqrt{\widetilde\Lambda}\left(ns\left(\eta,\frac{1}{2}\right)+sn\left(\eta,\frac{1}{2}\right)\right)}
  \notag\end{eqnarray}
\begin{eqnarray}
\qquad&=& \frac{2sd\left(\eta,\frac{1}{2}\right)}{\sqrt{\widetilde\Lambda}\sqrt{-\frac{3}{4}sd^4\left(\eta,\frac{1}{2}\right)-2sd^2\left(\eta,\frac{1}{2}\right)+4}}
 \notag\\
&&\notag\\
\qquad\qquad&=&\frac{\sqrt{3sc^4\left(\eta,\frac{1}{2}\right)+7sc^2\left(\eta,\frac{1}{2}\right)+4}}{2\sqrt{\widetilde\Lambda}\left(cs\left(\eta,\frac{1}{2}\right)+sc\left(\eta,\frac{1}{2}\right)\right)}\notag
\end{eqnarray}
since $\gamma^2=e_1-e_3=1$ and $k^2=\frac{e_2-e_3}{e_1-e_3}=1/4$.  In terms of theta functions,
\begin{eqnarray}
&&\\
a(\eta)&=& \frac{\theta_3\theta_1(\eta/\theta_3^2) \sqrt{\theta_3^4\theta_1^4(\eta/\theta_3^2)-5\theta_2^2\theta_3^2\theta_1^2(\eta/\theta_3^2)\theta_4^2(\eta/\theta_3^2)+4\theta_2^4\theta_4^4(\eta/\theta_3^2)}   }{ 2\sqrt{\widetilde\Lambda} \ \left( \theta_2^2\theta_4^2(\eta/\theta_3^2)+\theta_3^2\theta_1^2(\eta/\theta_3^2)   \right)  }
  \notag\\
&&\notag\\
&=&  \frac{2\theta_2\theta_4\theta_3^2 \theta_1(\eta/\theta_3^2)\theta_3(\eta/\theta_3^2)}{\sqrt{\widetilde\Lambda}\sqrt{-\frac{3}{4}\theta_3^8\theta_1^4(\eta/\theta_3^2)-2\theta_2^2\theta_4^2\theta_3^4\theta_1^2(\eta/\theta_3^2)\theta_3^2(\eta/\theta_3^2)+4\theta_2^4\theta_4^4\theta_3^4(\eta/\theta_3^2)}}
  \notag\\
&&\notag\\
&=&\frac{\theta_3\theta_1(\eta/\theta_3^2)\sqrt{3\theta_3^4\theta_1^4(\eta/\theta_3^2)+7\theta_3^2\theta_4^2\theta_1^2(\eta/\theta_3^2)\theta_2^2(\eta/\theta_3^2)+4\theta_4^4\theta_2^4(\eta/\theta_3^2)}}{2\sqrt{\widetilde\Lambda} \ \theta_4\theta_2(\eta/\theta_3^2)  \left( \theta_4^2\theta_2^2(\eta/\theta_3^2)+\theta_3^2\theta_1^2(\eta/\theta_3^2)\right)}\notag
\end{eqnarray}
where the suppressed $\tau$ is such that $1/4=\theta_2^4/\theta_3^4$.


\newpage
\section{\protect Summary}
\indent\indent There are a number of ways to see that elliptic and theta functions solve the $(d+1)-$dimensional Einstein gravitational field equations in a FRLW cosmology with a cosmological constant.  Here we considered a scenario with no scalar field and with density functions $\rho(t)=D_1/\widetilde{a}(t)^3+D_2/\widetilde{a}(t)^4$ for $D_1,D_2>0$, $\rho(t)=D/\widetilde{a}(t)^3$ and $\rho(t)=D/\widetilde{a}(t)^4$ scaling in inverse proportion to the scale factor $\widetilde{a}(t)$.  In these cases the first Einstein equation (i) takes the form $\dot{\widetilde{a}}(t)^2=$ an expression containing negative powers of the cosmic scale factor $\widetilde{a}(t)$.  At this point, one could have introduced the inverse function $y(x)$ of $\widetilde{a}(t)$ to obtain an expression for $y(x)$ as the integral of a power of $x$ divided by the square root of a polynomial in $x$.  That is, $y(x)$ would be an elliptic integral that is not normal; other authors have taken this approach \cite{ACB, KW}.  Here, we switched to conformal time by a change of variables $a(\eta)\stackrel{def.}{=}\widetilde{a}(f(\eta))$.  This produced an equation of the form $a'(\eta)^2=$ an expression containing nonnegative powers of the conformal scale factor $a(\eta)$.  

After reviewing the definitions and properties of elliptic and theta functions in sections 2 and 3, we introduced the FRLW cosmological model in section 4.  In section 5 for $\rho(t)=D/\widetilde{a}(t)^3$, we obtained a differential equation for $a(\eta)$ containing only even powers of $a(\eta)$ and constructed solutions in terms of Jacobi elliptic functions, restricted to particular values of the constant $D$, parameterized by modulus $0<k<1$.  The equivalent theta function representations for these solutions were recorded, and we noted the special cases for which the elliptic solutions reduce to elementary functions and the corresponding solution in cosmic time was also computed.  In section 6, we considered each of $\rho(t)=D_1/\widetilde{a}(t)^3+D_2/\widetilde{a}(t)^4$, $\rho(t)=D/\widetilde{a}(t)^3$ and $\rho(t)=D/\widetilde{a}(t)^4$ with various initial conditions and obtained solutions in terms of Weierstrass functions for general curvature $k'$ and constants $D_1,D_2,D>0$.  By considering these solutions  restricted to certain values of $D>0$ for $\rho(t)=D/\widetilde{a}(t)^3$ and $\rho(t)=D/\widetilde{a}(t)^4$, we wrote $a(\eta)$ in these cases equivalently in terms of both Jacobi elliptic and theta functions.

In current joint work with Floyd Williams \cite{JDFW2}, it is shown that elliptic functions also appear in the presence of a scalar field $\phi(t)$, for both the FRLW and Bianchi I $d-$dimensional cosmological models with a nonzero cosmological constant  and with a similar density function $\rho(t)$ scaling in inverse proportion to the cosmic scale factor $\widetilde{a}(t)$.  There we note that the equations of each of these cosmological models can be rewritten in terms of a generalized Ermakov-Milne-Pinney differential equation \cite{LidseyBEC, JDFW, JD1}, a type which the square root of the second moment of the wave function of the Bose-Einstein condensate (BEC) also satisfies.  From this we establish a direct mapping between cosmological and Bose-Einstein quantities.  On the cosmological side of the FRLW-BEC correspondence, by imposing an equation of state $\rho_\phi(t)=wp_\phi(t)$ ($w$ constant) on the density $\rho_\phi(t)$ and pressure $p_\phi(t)$ of the scalar field $\phi(t)$ one can obtain a differential equation involving elliptic functions on the BEC side of the mapping.

\end{document}